\documentclass[10pt,a4paper]{article}
\usepackage[T1]{fontenc}
\usepackage{amsmath,amsfonts, amssymb, amsbsy, mathrsfs, mathtools, bbm, enumitem}
\usepackage{titlesec}
\usepackage{float}
\usepackage{makeidx}
\usepackage{graphicx}
\usepackage{color}
\usepackage{url}
\usepackage{authblk}
\usepackage{multirow}
\usepackage{lscape}

\usepackage{mathpazo}
\usepackage[toc]{appendix}
\usepackage[hidelinks,colorlinks=true,linkcolor=red,citecolor=blue,urlcolor=magenta]{hyperref}
\usepackage[left=2.50cm, right=2.50cm, top=3.0cm, bottom=3.00cm]{geometry}

\usepackage{fancyhdr}

\newcommand{\email}[1]{\href{mailto:#1}{#1}}

\newcommand{\df}{\textrm{d}}

\newcommand{\cg}{\textnormal{\textsl{g}}}
\newcommand{\dprime}{{\prime\prime}}

\newcommand{\ex}{\textrm{e}}
\newcommand{\hf}{{\frac{1}{2}}}
\newcommand{\un}[1]{\underline{#1}}

\newcommand{\la}{\left\langle}
\newcommand{\ra}{\right\rangle}

\renewcommand{\i}{{\rm i}}

\newcommand{\bz}{\boldsymbol{Z}}

\newcommand{\B}{\mathcal{B}}

\renewcommand{\bar}{\overline}

\titleformat{\paragraph}
{\normalfont\normalsize\bfseries}{\theparagraph}{1em}{}

\titleformat{\subparagraph}
{\normalfont\normalsize\bfseries}{\thesubparagraph}{1em}{}

\usepackage[square,numbers,sort&compress]{natbib}

\numberwithin{equation}{section}

\allowdisplaybreaks[1]

\begin{document}
	\setlength{\bibsep}{0pt}
	
	\title{\textbf{Spherical and Plane Symmetric Solutions in Macroscopic Gravity}}
	\author{Anish Agashe\thanks{\email{anish.agashe@utdallas.edu}}\ }
	\author{Mustapha Ishak\thanks{\email{mishak@utdallas.edu}}}
	\affil{\small Department of Physics, The University of Texas at Dallas,\\800 W Campbell Rd, Richardson, TX, 75080, USA}
	
	\maketitle
	
	\begin{abstract}
		The theory of macroscopic gravity provides a formalism to average the Einstein field equations from small scales to largest scales in space-time. It is well known that averaging is an operation that does not commute with calculating the Einstein tensor and this leads to a correction term in the field equations known as back-reaction. In this work, we derive exact solutions to the macroscopic gravity field equations assuming that the averaged geometry is plane or spherically symmetric, and the source is taken as vacuum, dust, or perfect fluid. We then focus on the specific cases of spherical symmetry and derive solutions that are analogous to the Schwarzschild, Tolman VII, and Lema\^{\i}tre-Tolman-Bondi solutions. The geodesic equations and curvature structure are contrasted with the general relativistic counterparts for the Schwarzschild and Lema\^{\i}tre-Tolman-Bondi solutions.  
	\end{abstract}

	%
	
	\section{Introduction}	
	An outstanding problem in general relativity (GR) is how the small scale lumpy universe is averaged to be homogeneous at the largest scales. This problem is not easily reconcilable due to the difficulties in defining a mathematically rigorous averaging procedure for non-trivial geometries such as the pseudo-Riemannian geometry of general relativity. Moreover, the Einstein tensor is a non-linear function of the metric which would mean that the average of the Einstein tensor for a given metric is not equal to the Einstein tensor constructed with the average of that metric, i.e., $ \la \boldsymbol{E}[\textsl{\textbf{g}}] \ra \ne \boldsymbol{E}[\la\textsl{\textbf{g}}\ra] $. These issues are collectively known as the averaging problem in general relativity \cite{ellis2,ellis3,ellis4,shiro1,shiro2,shiro3,tava,cliftonbr,buchertbr,ellisbr,ellisbr1,hoogen1}.
	
	Therefore, when dealing with large length scales, a set of averaged equations should be used. In these equations, a correction term should be present to account for the non-commutativity between averaging and calculating the Einstein tensor \cite{ellis2,ellis3,ellis4,shiro1,shiro2,shiro3}. This term encapsulates the effect of the microscopic structure of the space-time on the dynamics of the macroscopic one, and is sometimes referred to as back-reaction. Macroscopic gravity (MG) is a formalism for covariant space-time averaging \cite{zala1,zala2,zala3,zala4,zala5,zala6,zala7} which offers a promising solution to the averaging problem. It is an exact (non-perturbative) approach which employs averaging bivectors and Lie dragging of the averaging regions. It is valid for arbitrary classical tensor fields on a general $n$-dimensional differentiable manifold.
	
	The field equations of MG are the macroscopic analogue of the Einstein field equations (EFEs). They are characterised by a tensorial correction term which is a combination of various traces of a quantity called connection correlation. They can be written in the form of EFEs by taking this correction term to the right hand side of the equations. Then, the dynamics of a (assumed) macroscopic space-time geometry can be analysed by determining this correction term. This can be done by imposing some reasonable assumptions on the connection correlation \cite{coley,hoogen2,clifton}. The solutions to the averaged field equations in MG, with the macroscopic geometry assumed to be FLRW, have been presented in \cite{coley,hoogen2,clifton,tharake1}. Other exact macroscopic geometries like Bianchi Type-I \cite{tharake1} and static spherically symmetric (Schwarzschild) \cite{hoogenss} have also been explored. Linear perturbations around the FLRW geometry have been analysed in \cite{clifton}. In \cite{coleyss1,coleyss2,tegai}, the authors took the microscopic geometry to be spherically symmetric and wrote it in volume preserving coordinates. Then, they averaged it using Zalaletdinov's procedure which becomes trivial in the volume preserving coordinates, since the averaging operators reduce to Kronecker delta \cite{zala3}.
	
	
	In this paper, we derive solutions to the MG field equations where the macroscopic geometry is assumed to be spherical or plane symmetric (that is, the less symmetric microscopic geometry averages out to be spherical/plane symmetric). We start by assuming that the macroscopic geometry admits the $G_3$ group of motion (on $V_2$) which contains two special cases -- plane and spherical symmetry\footnote{See chapters 15 and 16 in \cite{stephani} and the references within for the details.} \cite{stephani}, and then, calculate the MG correction term for such geometries. Our approach to calculate the correction term is similar to the one taken in \cite{hoogenss} to study static spherically symmetric solutions. Here, we extend the treatment in \cite{hoogenss} to include non-static spherically symmetric and plane symmetric geometries as well. We find that the MG correction term takes a form of an anisotropic fluid with a qualitative behaviour of an effective spatial curvature in the field equations. Recently, such geometries were also analysed in the context of other averaging formalisms \cite{clifton1}.
	
	We categorise the solutions based on the source for the space-time -- vacuum, dust, and perfect fluid. Within these three categories, we treat, in detail, the cases of the static spherically symmetric vacuum solution (Schwarzschild exterior \cite{schw1}), the static spherically symmetric perfect fluid solutions (Schwarzschild interior \cite{schw1} and Tolman VII \cite{tolman1,wyman,mehra,durgapal1,durgapal2,delgaty,neary1,neary2,raghoo,vaidya}), and the non-static spherically symmetric dust solution (Lema\^itre-Tolman-Bondi (LTB) \cite{lemaitre,lemaitre1,tolman,bondi,datt1,datt2}). These solutions are the exact solutions to the MG field equations and do not represent the same geometry as their counterparts in GR, since they are not the solutions of Einstein field equations. This is in the same spirit as that of GR -- the matter distribution (at the macroscopic level) should determine the (macroscopic) geometry through the macroscopic field equations. 
	
	The paper is arranged in the following manner: In section \ref{sec-mgtheory}, we briefly review the macroscopic gravity formalism. In section \ref{sec-inhomsol}, we present the spherical and plane symmetric solutions to the MG field equations. Then, in sections \ref{subsec-vacuum}, \ref{subsec-dust} and \ref{subsec-perfectfluid}, we analyse specific solutions by assuming the source to be a vacuum, dust and perfect fluid, respectively. The MG analogue of the Schwarzschild, Tolman VII and LTB solutions are derived. Finally, we discuss the results in this paper and make some general remarks in section \ref{sec-discussion}. 
	
	The notation and the convention used are as follows: objects associated with the microscopic geometry are denoted by lowercase letters and those with the macroscopic geometry by uppercase letters. Greek indices run from $0$  to $3$ and Latin indices from $1$ to $3$. Angular brackets $ \la\cdots\ra $ denote the averaging operation or sometimes averaged quantities. Covariant differentiation with respect to the macroscopic connection is denoted by $||$. Indices with round brackets $ (~) $/square brackets $[~]$ are symmetrised/anti-symmetrised; and  underlined indices are not included in (anti-)symmetrisation. The covariant derivative is denoted by $\nabla_\mu$. The sign convention followed is the `Landau-Lifshitz Space-like Convention (LLSC)' \cite{mtw}. That is, the signature of the metric is taken to be Lorentzian $ (-,+,+,+) $, the Riemann curvature tensor is defined as, $ {r^\mu}_{\alpha\nu\beta}~=~2\partial_{[\nu}{\gamma^\mu}_{\underline{\alpha}\beta]} + 2{\gamma^\epsilon}_{\alpha[\beta}{\gamma^\mu}_{\underline{\epsilon}\nu]} $ and $ r_{\mu\nu}~=~{r^\alpha}_{\mu\alpha\nu} $ is the Ricci tensor. The Ricci scalar is defined as $ r~=~{r^\mu}_\mu~=~\cg^{\mu\nu}r_{\mu\nu} $.  Finally, the units are taken such that $ G=1=c $, i.e., $ \kappa = 8\pi $. 
	
	\section{Macroscopic Gravity Field Equations and their Solutions} \label{sec-mgtheory}
	Using the concepts of macroscopic electrodynamics \cite{russ,lorentz,jackson}, a covariant averaging procedure was developed by Zalaletdinov \cite{zala1,zala2,zala3,zala4,zala5,zala6,zala7}, which can be used in general relativity. It is a generalisation of averaging on Minkowski space-time and is based on the concept of Lie dragging of the averaging regions. This procedure is valid for arbitrary classical tensor fields on any differentiable manifold \cite{zala1,zala2,zala3,zala4,zala5,zala6,zala7}.
	
	Let there be a geometric object (vector, tensor etc.) $ p^\alpha_\beta(x) $ defined on an $ n $-dimensional differentiable metric manifold $ (\mathcal{M}, \cg_{\alpha\beta}) $. Then, the space-time averaged value of this object over a compact region $ \Sigma \subset \mathcal{M} $ with a volume $ n $-form around a supporting point $ x \in \Sigma $, is defined as,
	\begin{equation} \label{mgavg}
		\la p^\alpha_\beta(x) \ra = \dfrac{\int_\Sigma \mathcal{A}^\alpha_{\mu^\prime}(x,x^\prime) p^{\mu^\prime}_{\nu^\prime}(x^\prime) \mathcal{A}^{\nu^\prime}_\beta(x^\prime, x) \sqrt{-\cg^\prime}\ \df^n x^\prime}{\int_\Sigma \sqrt{-\cg^\prime}\ \df^n x^\prime}
	\end{equation}
	where $ \int_\Sigma \sqrt{-\cg^\prime}\ \df^n x^\prime $ is the volume ($ V_\Sigma $) of the region $ \Sigma $ and $ \cg^\prime = \det\left[\cg_{\alpha\beta}(x^\prime)\right] $. The integration is done over all the points $ x^\prime \in \Sigma $. The integrand $ \mathcal{A}^\alpha_{\mu^\prime}(x,x^\prime) p^{\mu^\prime}_{\nu^\prime}(x^\prime) \mathcal{A}^{\nu^\prime}_\beta(x^\prime, x) $ is called the \emph{bilocal extension} of the object $ p^\alpha_\beta(x) $; and the objects $ \mathcal{A}^\alpha_{\mu^\prime}(x,x^\prime) $ and $ \mathcal{A}^{\nu^\prime}_\beta(x^\prime, x) $ are called the \emph{bilocal averaging operators}.
	
	Then, bilocal objects $\boldsymbol{\mathcal{F}}$ are defined in terms of these averaging operators as,
	\begin{equation}\label{bilextconncoeff}
		{\mathcal{F}^\alpha}_{\beta\rho} = \mathcal{A}^\alpha_{\epsilon^\prime}\left(\partial_\rho\mathcal{A}^{\epsilon^\prime}_\beta + \nabla_{\sigma^\prime}\mathcal{A}^{\epsilon^\prime}_\beta\mathcal{A}^{\sigma^\prime}_\rho\right)
	\end{equation} 
	These objects behave as connection coefficients at $ x $ and hence, can be considered as the bilocal extension of the microscopic connection coefficients. The averages of these objects, $ \la{\mathcal{F}^\alpha}_{\beta\rho}\ra $ serve as the affine connection coefficients of the averaged space-time. 
	
	Further, the average of the microscopic Riemann curvature tensor, $ \la{r^\alpha}_{\beta\rho\sigma}\ra $ is written as $ {R^\alpha}_{\beta\rho\sigma} $. A curvature tensor corresponding to the macroscopic connection can be calculated, and turns out to be related to the average of the microscopic Riemann curvature tensor through the following formula,
	\begin{equation}\label{macrocurv}
		{M^\alpha}_{\beta\rho\sigma} = {R^\alpha}_{\beta\rho\sigma} + 2\la{\mathcal{F}^\delta}_{\beta[\rho}{\mathcal{F}^\alpha}_{\underline{\delta}\sigma]}\ra - 2\la{\mathcal{F}^\delta}_{\beta[\rho}\ra\la{\mathcal{F}^\alpha}_{\underline{\delta}\sigma]}\ra
	\end{equation}
	
	Then, in order to define splitting rules for averaging the differential Bianchi identities, correlation tensors need to be defined. The connection correlation is a six rank tensor which takes into account the non-commutativity of averaging and calculating the Einstein tensor. It is defined as,
	\begin{equation}\label{corr2form}
		{Z^\alpha}_{\beta[\gamma}{^\mu}_{\underline{\nu}\sigma]} = \la {\mathcal{F}^\alpha}_{\beta[\gamma}{\mathcal{F}^\mu}_{\underline{\nu}\sigma]} \ra - \la {\mathcal{F}^\alpha}_{\beta[\gamma}\ra \la{\mathcal{F}^\mu}_{\underline{\nu}\sigma]} \ra
	\end{equation}
	
	Note that the second term on the right hand side in equation \eqref{macrocurv} is merely a contraction of the connection correlation and can be written as, $ {Q^\alpha}_{\beta\rho\sigma}~=~2\la{\mathcal{F}^\delta}_{\beta[\rho}{\mathcal{F}^\alpha}_{\underline{\delta}\sigma]}\ra - 2\la{\mathcal{F}^\delta}_{\beta[\rho}\ra\la{\mathcal{F}^\alpha}_{\underline{\delta}\sigma]}\ra=2{Z^\epsilon}_{\beta[\rho}{^\alpha}_{\underline{\epsilon}\sigma]}$. This tensor is like a curvature tensor in its mathematical construction and follows the algebraic properties of one. It can be considered as a `curvature deformation' tensor since it measures the difference between the macroscopic curvature tensor and the average of the microscopic curvature tensor. Higher order correlations also exist (restricted up to the dimensionality of the macroscopic space-time), but can be taken to be zero in a self consistent manner \cite{zala1,zala2}. This makes the formalism more practical with equations involving only the connection correlation. 
	
	The final form of the averaged Einstein field equations is given by,
	\begin{equation}\label{macroefe}
		{E^\epsilon}_\gamma = G^{\beta\epsilon} M_{\beta\gamma} - \hf\delta^\epsilon_\gamma G^{\mu\nu} M_{\mu\nu} = \kappa {T^\epsilon}_\gamma + \left({Z^\epsilon}_{\mu\nu\gamma} - \hf\delta^\epsilon_\gamma Q_{\mu\nu}\right) G^{\mu\nu} 
	\end{equation}
	where, $ {T^\epsilon}_\gamma = \la {t^\epsilon}_\gamma \ra $ is the average of the energy-momentum tensor for the microscopic matter distribution, $G^{\mu\nu}$ is the inverse of the macroscopic metric tensor, $ M_{\beta\gamma} $ is the macroscopic Ricci tensor, $ {Z^\epsilon}_{\mu\nu\gamma}~=~2{Z^\epsilon}_{\mu[\alpha}{^\alpha}_{\underline{\nu}\gamma]} $ is a Ricci-tensor-like object for the connection correlation and $ Q_{\mu\nu}~=~{Q^\epsilon}_{\mu\epsilon\nu} $.
	
	Therefore, in the theory of macroscopic gravity, averaging out the Einstein field equations introduces additional terms on right hand side of these equations, constituting various traces (contractions) of the connection correlation. We call these additional terms as the MG correction term, denoted as,
	\begin{equation}\label{tmg}
		{C^\epsilon}_\gamma = \left({Z^\epsilon}_{\mu\nu\gamma} - \frac{1}{2}\delta^\epsilon_\gamma Q_{\mu\nu}\right) G^{\mu\nu}
	\end{equation}
	
	To derive a solution to the averaged field equations \eqref{macroefe}, we need to determine the form of this correction term explicitly for a given macroscopic geometry. This is done by virtue of several algebraic and differential constraints on the connection correlation \cite{zala1,zala2},
	\begin{subequations}
		\begin{align}
			{Z^\alpha}_{\beta\gamma}{^\mu}_{\nu\sigma} &=  -{Z^\alpha}_{\beta\sigma}{^\mu}_{\nu\gamma} \label{corr2formprop1}\\
			{Z^\alpha}_{\beta\gamma}{^\mu}_{\nu\sigma} &=  -{Z^\mu}_{\nu\gamma}{^\alpha}_{\beta\sigma} \label{corr2formprop2}	\\	
			{Z^\alpha}_{\beta[\gamma}{^\mu}_{\nu\sigma]} &= 0 \label{corr2formprop3}\\
			{Z^\epsilon}_{\epsilon\gamma}{^\mu}_{\nu\sigma} &= 0 \label{corr2formprop4}\\
			{{{Z^\alpha}_{\beta\mu}}^\gamma}_{\delta\nu}u^\nu &= 0 \label{corr2formprop5}\\
			{Z^\alpha}_{\beta[\gamma}{^\mu}_{\underline{\nu}\sigma||\lambda]} &= 0 \label{corr2formprop6}
		\end{align}
		\begin{equation}\label{corr2formprop7}
			{Z^\epsilon}_{\beta[\gamma}{^\mu}_{\underline{\nu}\sigma}{M^\alpha}_{\underline{\epsilon}\lambda\rho]} - {Z^\alpha}_{\epsilon[\gamma}{^\mu}_{\underline{\nu}\sigma}{M^\epsilon}_{\underline{\beta}\lambda\rho]} + {Z^\alpha}_{\beta[\gamma}{^\epsilon}_{\underline{\nu}\sigma}{M^\mu}_{\underline{\epsilon}\lambda\rho]} - {Z^\alpha}_{\beta[\gamma}{^\mu}_{\underline{\epsilon}\sigma}{M^\epsilon}_{\underline{\nu}\lambda\rho]} = 0 
		\end{equation}

	\end{subequations}
	where, $\boldsymbol{u}$ is the unit time-like 4-vector field and $\boldsymbol{M}$ is the Riemann curvature tensor for the (assumed) macroscopic space-time geometry.   
	
	Equations \eqref{corr2formprop1}-\eqref{corr2formprop5} are just algebraic constraints and do not depend on the macroscopic geometry. Solving these leaves us with 121 independent components in the connection correlation. Equations \eqref{corr2formprop5}, \eqref{corr2formprop6} and \eqref{corr2formprop7} ensure that the higher order correlations are zero. Then, all one has to deal with is the connection correlation. There are several assumptions on the structure of the macroscopic geometry and the functional form of the connection correlation that go into solving the latter two of these \cite{hoogenss}. Given these assumptions, the connection correlation can be determined completely. A systematic way of solving the MG equations has been presented in \cite{zala4,hoogenss,hoogen2,tharake1}.
	
	In the following sections, starting with the 121 components, we solve the last two equations to determine the MG correction term for spherical and plane symmetric macroscopic geometries. 
	
	\section{Spherical and Plane Symmetric Solutions} \label{sec-inhomsol}
	
	For a coordinate system $(x^0,x^1,x^2,x^3)$, the general form of the line element for a spherical and plane symmetric space-time is given by \cite{stephani,goenner},
	\begin{equation}\label{g3metric}
		\df s^2 = G_{\mu\nu}\df x^\mu \df x^\nu = -\ex^{2 P}(\df x^0)^2 + \ex^{2 Q}(\df x^1)^2 + R^2\left[ (\df x^2)^2 + S^2(x^2,k)(\df x^3)^2 \right]
	\end{equation}
	where, $G_{\mu\nu}$ is the macroscopic metric, $P,Q,R$ are, in general, arbitrary functions of coordinates $(x^0,x^1)$, and just the coordinate $x^1$ for the static case. The time-like unit 4-vector field (the 4-velocity of time-like particles) admitted by such space-time is given by, $u^\nu = \left[\ex^{-P},0,0,0\right] $. The function $S(x^2,k)$ takes the values $\{\sin (x^2),1\}$ for $k=\{1,0\}$, where $k~=~KR^2$ and $K$ is the constant Gaussian curvature of the 2-subspace ($ x^0,x^1 = $ const.). Since the function, $R$, has a fixed value on this 2-subspace, one can normalise the curvature in this manner \cite{stephani}. The spherical and plane symmetric cases are represented by $k=1$ and $k=0$, respectively. 
	
	It is conventional to use spherical coordinates $(t,r,\theta,\phi)$ in the case of spherical symmetry $ (k=1) $. Then, the line element takes the form,
	\begin{equation}\label{ssmetric}
		\df s^2 = -\ex^{2 P}\df t^2 + \ex^{2Q}\df r^2 + R^2 \left( \df \theta^2 + \sin^2\theta\df\phi^2 \right)
	\end{equation}
	where $P,Q,R$ are arbitrary functions of time, $t$, and the radial coordinate, $r$, for the non-static case. In the static case, $P,Q$ are functions of only $r$ and the function $R$ can be chosen to be equal to $r$ \cite{stephani}.
	
	For the case of plane symmetry $(k=0)$, we often use Cartesian coordinates $(t,x,y,z)$. Then, the line element takes the form,
	\begin{equation}\label{psmetric}
		\df s^2 = -\ex^{2 P}\df t^2 + \ex^{2Q}\df x^2 +  R^2\left( \df y^2 + \df z^2  \right)
	\end{equation}
	where $P,Q,R$ are arbitrary functions of time, $t$, and the spatial coordinate, $x$, for the non-static case\footnote{Note that, one can choose any one of the three spatial coordinates to be $ x^1 $. Then, the metric will be modified accordingly and so would the MG correction term.}. In the static case, $P,Q,R$ are functions of only $x$.
	
	We calculate the MG correction term with the line element for the macroscopic geometry to be the one in equation \eqref{g3metric}\footnote{We did the same calculations separately for equations \eqref{ssmetric} and \eqref{psmetric} for both static and non-static cases. The results were identical to the respective cases presented here.}. We take the components of connection correlation to be dependent on all the coordinates, i.e., $\bz~\equiv\bz(x^0,x^1,x^2,x^3)$. We start with the 121 independent components in $\bz$ that remain after solving constraints \eqref{corr2formprop1}-\eqref{corr2formprop5}. Solving the differential constraint \eqref{corr2formprop6} leaves only 7 independent components. The integrability condition \eqref{corr2formprop7} is trivially satisfied and does not reduce the independent components any further.
	
	These components are characterised by arbitrary functions of coordinates $x^2$ and $x^3$. To make the MG correction term compatible with the macroscopic geometry, two of these need to be constant. Then, out of the 7 independent components in $\bz$, five are arbitrary functions and two are constants. We label these as $f_n(x^2,x^3),\ (n=1,...,5)$ and $b_1,b_2$. The number of non-zero components in the connection correlation tensor is 44 and that in the curvature deformation tensor, $\boldsymbol{Q}$, is 8. The non-Riemannian curvature tensor, $\boldsymbol{R}$, has 44 non-zero components comprising of three out of the seven variables in $\bz$. We do not list them here as the real important quantity is the MG correction term, since it enters the field equations. The MG correction term is diagonal and takes the following form,
	\begin{subequations}\label{statsphersym}
		\begin{align}
			{C^0}_0 &= \frac{\B}{R^2} \\
			{C^0}_i &= 0 = {C^i}_0 \\
			{C^1}_1 &= \frac{\B}{R^2}\; ; \quad {C^2}_2 = 0 = {C^3}_3
		\end{align}
	\end{subequations}
	Writing in matrix form, this looks like,
	\begin{equation}\label{statsphersymmatrix}
		{C^\epsilon}_\gamma = \frac{\B}{R^2}\begin{bmatrix}
			&1 &0 &0 &0&\\
			&0 &1 &0 &0&\\
			&0 &0 &0 &0&\\
			&0 &0 &0 &0&
		\end{bmatrix}
	\end{equation}	
	where, $ \B = 2(b_1+b_2) $ is a constant.   
	
	The MG correction term takes the form of an anisotropic fluid. Only the radial component of the effective pressure is non-zero and is equal in magnitude, but opposite in sign, to the effective energy density. The MG correction term for the static spherically symmetric geometry in our case is the same as what was found in \cite{hoogenss}, although we have applied fewer restrictions on $\bz$.
	
	\subsection{The Macroscopic Einstein Field Equations}
	We take the average of the microscopic energy-momentum tensor to be that of an inhomogeneous relativistic perfect fluid with an isotropic pressure,
	\begin{equation}\label{emtensor}
		{T^\epsilon}_\gamma  =  (\rho + p) u^\epsilon u_\gamma + p\delta^\epsilon_\gamma	
	\end{equation}
	where, $\rho\equiv\rho(x^0,x^1)$ and $p\equiv p(x^0,x^1)$ are the energy density and pressure. Taking the averaged matter distribution to be a perfect fluid does not reduce the generality of the equations that follow (for the purposes of this paper), since both vacuum and dust can be considered as special cases. The conservation of energy-momentum\footnote{The differential constraint \eqref{corr2formprop6} of the formalism ensures that the MG correction term has also vanishing covariant derivative \cite{zala1,zala2}. Hence, the covariant divergence of the two sides of equation \eqref{macroefe} vanishes identically -- the left side due to the usual differential Bianchi identity.}, $\nabla_\epsilon {T^\epsilon}_\gamma = 0 $, gives two equations -- one each for $\gamma~=~0,1$, while other components are identically satisfied,
	\begin{subequations}\label{conseq}
		\begin{align}
			\frac{\dot{\rho}}{\rho + p} + \dot{Q} + 2\frac{\dot{R}}{R} &= 0 \label{conseq1}\\
			\frac{p^\prime}{\rho + p} + P^\prime &= 0 \label{conseq2}
		\end{align}
	\end{subequations} 
	
	Then, without assuming any equation of state, $p(\rho)$, the macroscopic field equations for the metric in \eqref{g3metric} read as 	\begin{subequations}\label{genmefe}
		\begin{equation}\label{mefe1}
			{E^0}_0 = \frac{\ex^{-2Q}}{R^2}\left[ 2RR^\dprime + R^\prime \left(R^\prime - 2RQ^\prime \right) \right] - \frac{\ex^{-2P}}{R^2} \left[ \dot{R} \left(\dot{R} + 2R\dot{Q}\right) \right] - \frac{k}{R^2} = -8\pi \rho + \frac{\B}{R^2} 
		\end{equation}
		\begin{equation}\label{mefe2}
			{E^0}_1 = \frac{2\ex^{-2P}}{R} \left[ \dot{R}^\prime - \dot{Q}R^\prime - P^\prime\dot{R} \right] = 0
		\end{equation}
		\begin{equation}\label{mefe3}
			{E^1}_0 = -\frac{2\ex^{-2Q}}{R} \left[ \dot{R}^\prime - \dot{Q}R^\prime - P^\prime\dot{R} \right] = 0
		\end{equation}
		\begin{equation}\label{mefe4}
			{E^1}_1 = - \frac{\ex^{-2P}}{R^2} \left[ 2R\ddot{R} + \dot{R}\left( \dot{R} - 2R\dot{P}\right) \right] + \frac{\ex^{-2Q}}{R^2}\left[ R^\prime \left(R^\prime + 2R P^\prime\right) \right] - \frac{k}{R^2} = 8\pi p + \frac{\B}{R^2} 
		\end{equation}
		\begin{multline}\label{mefe5}
			{E^2}_2 = {E^3}_3 = \frac{\ex^{-2Q}}{R}\left[ R^\dprime + R^\prime \left( P^\prime -Q^\prime\right) + R\left( P^\dprime + {P^\prime}^2 - P^\prime Q^\prime \right)\right]\\ - \frac{\ex^{-2P}}{R} \left[ \ddot{R} + \dot{R}\left( \dot{Q} - \dot{P}\right) + R\left(\ddot{Q} + {\dot{Q}}^2 - \dot{P}\dot{Q}\right) \right] = 8\pi p
		\end{multline}
	\end{subequations}
	where, `dot' and `prime' represent differentiation with respect to the coordinates $x^0$ and $x^1$, respectively. The plane and spherical symmetric cases correspond to $k=0$ and $k=1$, respectively. The MG correction term enters equations \eqref{mefe1} and \eqref{mefe4} as an additional curvature term. Other equations are not affected. In the case of static solutions, all the terms with a dot go to zero.
	
	\section{Vacuum} \label{subsec-vacuum}
	In the case of vacuum, the conditions from conservation of energy are identically satisfied. The MG correction term does not change the field equations qualitatively since the constant term $\B$ can be absorbed in the curvature term. Therefore, the usual results for the analysis of vacuum solutions hold. However, the terms with $k$ would be replaced by $k+\B$. Therefore, the generalised Birkhoff theorem \cite{stephani,birkhoff,cahen,barnes,goenner1,bona} would still hold and the metric would be necessarily static. This means that all the terms with `dot' vanish. Moreover, if $\partial_\mu R \partial^\mu R >0 $, one can choose canonical coordinates, where $R = x^1$ \cite{stephani,goenner}. Then, solving  equation \eqref{mefe1}, we would find, 
	\begin{equation}\label{vacsol1}
		\ex^{2Q} = \left[(k+\B) - \frac{2m}{x^1}\right]^{-1}
	\end{equation}
	where, $m$ is a constant that comes from integration and $Q$ is a function of only $x^1$ now. Further, substituting equation \eqref{mefe4} in \eqref{mefe1}, we get,
	\[ P^\prime = -Q^\prime \ \Rightarrow P = -Q + C_1 \]
	where $ C_1 $ is a constant of integration. This gives us,
	\begin{equation}\label{vacsol2}
		\ex^{2P} = C_2\ex^{-2Q} = C_2\left[ (k+\B) - \frac{2m}{x^1} \right]
	\end{equation}
	where, $C_2 = \ex^{2C_1}$. For simplicity, we can choose $C_2=1$, as one does in the usual  Schwarzschild solution. Then, using equations \eqref{vacsol1} and \eqref{vacsol2} in equation \eqref{g3metric}, the macroscopic metric takes the following form,
	\begin{equation}\label{vacmetric}
		\df s^2 = -f(x^1,k,\B) (\df x^0)^2 + \frac{1}{f(x^1,k,\B)} (\df x^1)^2 + (x^1)^2\left[ (\df x^2)^2 + S^2(x^2,k)(\df x^3)^2 \right]
	\end{equation}
	where $f = \left[ (k+\B) - \frac{2m}{x^1} \right]$. Therefore, the vacuum solution is now modified to have the back-reaction term in the line element. This metric has a coordinate singularity at $x^1 = \frac{2m}{(k+\B)}$.
	
	\subsection{The MG-Schwarzschild Exterior Solution}
	The static spherically symmetric case ($k=1$) gives us the MG-Schwarzschild (exterior) solution. The line element, in coordinates $(t,r,\theta,\phi)$, then, looks like,
	\begin{equation}\label{mgschw}
		\df s^2 = -\left[ (1+\B) - \frac{2m}{r} \right]\df t^2 + \frac{1}{\left[ (1+\B) - \frac{2m}{r} \right]} \df r^2 \\+ r^2\df \theta^2 + r^2 \sin^2 \theta \df \phi^2
	\end{equation}
	It reduces to the usual Schwarzschild exterior solution of GR when we take the back-reaction term $\B$ to be zero. The line element in equation \eqref{mgschw} clearly represents a geometry that is different from that of the Schwarzschild solution. We know that in the GR-Schwarzschild geometry, the Ricci tensor and hence the Ricci scalar is zero. But this is not the case in the MG-Schwarzschild geometry where, ${R^\mu}_\nu~=~{\rm diag}[0,0,-\frac{\B}{r^2},-\frac{\B}{r^2}]$ and $R~=~-\frac{2\B}{r^2}$. The Kretschmann scalar is also modified, and is given by, $K~=~\frac{48m^2}{r^6}~+~\frac{4\B r(\B r-4m)}{r^6}$. This is exactly the same result as in \cite{hoogenss}.
	
	As explained above, since the geometry is not Schwarzschild anymore, the geodesic equations will be modified as well. Using the line element in equation \eqref{mgschw}, the Lagrangian density will become,
	\begin{equation}\label{mgschwlag}
		2\mathcal{L} = -f(r,\B) \left(\frac{\df t}{\df \lambda}\right)^2 + \frac{1}{f(r,\B)} \left(\frac{\df r}{\df \lambda}\right)^2 + r^2\left(\frac{\df \phi}{\df \lambda}\right)^2 
	\end{equation}   
	where, $\lambda$ is an affine parameter and $f=\left(1+\B - \frac{2m}{r} \right)$. Using the Euler-Lagrange equations, we get, $\frac{\df t}{\df \lambda}~=~\frac{\gamma}{f(r,\B)};\ \frac{\df \phi}{\df \lambda}~=~\frac{l}{r^2}$, where, $\gamma$ and $l$ are constants. We know that for geodesics $2\mathcal{L}=\varepsilon$, where, $\varepsilon=-1,0,+1$ correspond to time-like, null and space-like geodesics, respectively. Using this, we can obtain the following orbital equation for the respective type of particles,
	\begin{equation}
		\left(\frac{\df u}{\df \phi}\right)^2 = 2mu^3 - (1+\B)u^2 - \frac{2m\varepsilon}{l^2}u + \frac{\gamma^2 + (1+\B)\varepsilon}{l^2} \label{mgschworbeq1}
	\end{equation}
	which gives,
	\begin{equation}
		\frac{\df^2 u}{\df \phi^2} = 3mu^2 -(1+\B)u - \frac{m\varepsilon}{l^2}
	\end{equation}
	where, $u\equiv\frac{1}{r}$. The above equation can be written in the form of the usual equation in GR by rescaling the coordinate $\phi$ and mass $m$ as $\bar{\phi}=\sqrt{1+\B}\phi$ and $\bar{m}=\frac{m}{1+\B}$, respectively. The solution to this equation is, then, given by,
	\begin{equation}\label{mgschworbeqsol}
		u = \frac{-m\varepsilon}{(1+\B)l^2}\left[ 1+\alpha \cos \left\{ \phi \left(\sqrt{1+\B} + \frac{3m^2\varepsilon}{(1+\B)^{\frac{3}{2}}l^2}\right) \right\} \right]
	\end{equation}
	where, $\alpha$ is a small parameter ($ |\alpha|<<1 $) that quantifies the deviation from the Newtonian solution.
	
	It is a fruitful exercise to write the metric in equation \eqref{mgschw} in various other coordinates that are frequently used to write the GR-Schwarzschild solution. This will be particularly useful when one wants to study the behaviour of various geodesics in the MG-Schwarzschild exterior space-time. In `isotropic' coordinates \cite{eddington}, we have,
	\begin{equation}\label{mgschwiso}
		\df s^2 = -(1+\B)\left(\frac{1-\frac{m}{2(1+\B)\bar{r}^{\sqrt{1+\B}}}}{1 + \frac{m}{2(1+\B)\bar{r}^{\sqrt{1+\B}}}} \right)^2\df t^2 + \bar{r}^{2\sqrt{1+\B} -2} \left( 1 + \frac{m}{2(1+\B)\bar{r}^{\sqrt{1+\B}}} \right)^4 \left(\df \bar{x}^2 + \df \bar{y}^2 + \df \bar{z}^2 \right)
	\end{equation}
	where, $ r = \bar{r}^{\sqrt{1+\B}}\left( 1 + \frac{m}{2(1+\B)\bar{r}^{\sqrt{1+\B}}} \right)^2 $ and we have, $\bar{x}~=~\bar{r} \sin \theta \cos \phi$; $\bar{y}~=~\bar{r} \sin \theta \cos \phi$; $\bar{z}~=~\bar{r}\cos\theta $. In the Eddington-Finkelstein coordinates \cite{eddington1,fink,pen}, we have,
	\begin{equation}\label{mgschwedfink1}
		\df s^2 = -\left[(1+\B) - \frac{2m}{r}\right]\df v^2 + 2 \df v \df r + r^2\df \theta^2 + r^2 \sin^2 \theta \df \phi^2
	\end{equation}
	where, $ v = t + \frac{r}{1+\B} + \frac{2m}{(1+\B)^2} \ln \left|\frac{(1+\B)r}{2m} - 1\right|$. Similarly, we can write,
	\begin{equation}\label{mgschwedfink2}
		\df s^2 = -\left[(1+\B) - \frac{2m}{r}\right]\df u^2 - 2 \df u \df r + r^2\df \theta^2 + r^2 \sin^2 \theta \df \phi^2
	\end{equation}
	where, $ u = t - \frac{r}{1+\B} - \frac{2m}{(1+\B)^2} \ln \left|\frac{(1+\B)r}{2m} - 1\right|$. In Kruskal-Szekeres \cite{kruskal,szekeresgm} coordinates, we have,
	\begin{equation}\label{mgschwks1}
		\df s^2 = -\frac{32 m^3}{r(1+\B)^4}\ex^{\frac{-(1+\B)r}{2m}} \df U\df V + r^2\df \theta^2 + r^2 \sin^2 \theta \df \phi^2 
	\end{equation}
	where, $V=\ex^{\frac{v(1+\B)^2}{4m}}$ and $U=-\ex^{\frac{-u(1+\B)^2}{4m}}$. This can be further written as,
	\begin{equation}\label{mgschwks2}
		\df s^2 = -\frac{32 m^3}{r(1+\B)^4}\ex^{\frac{-(1+\B)r}{2m}} (\df T^2 - \df X^2) + r^2\df \theta^2 + r^2 \sin^2 \theta \df \phi^2 
	\end{equation}
	where, $T=\hf(V+U)$ and $X=\hf(V-U)$.
	
	It should be possible to also write an equivalent form of the MG-Schwarzschild metric in coordinate systems like that of Lema\^itre-Novikov \cite{lemaitre,novikov} and Israel \cite{israel}. 
	
	\section{Dust} \label{subsec-dust}
	In the case of dust, the pressure, $p$, in equations \eqref{conseq} and \eqref{genmefe} is zero. Then, equation \eqref{conseq2} means that $P^\prime = 0$. This implies that $P\equiv P(x^0)$. We can divide the solutions into two classes: i) when $R^\prime=0$; $\dot{R}\ne 0$ and, ii) when $R^\prime \ne 0$. 
	
	\subsection{Solutions with $R^\prime = 0,\ \dot{R} \ne 0$}
	\un{\textit{When \emph{$R^\prime = 0,\ \dot{R} \ne 0$}}}, one can choose $R=x^0$ \cite{stephani} and then integrate equation \eqref{mefe4} to get,
	\begin{equation}\label{pdefdustcls1}
		\ex^{2P} = \left\{ \frac{C_1}{x^0} - (k+\B) \right\}^{-1} = \frac{x^0}{C_1 - (k+\B)x^0}
	\end{equation} 
	where, $C_1$ is a constant of integration.
	
	Then, assuming $\ex^Q=U(x^0,x^1)\ex^{-P}$, equations \eqref{mefe5} can be written as,
	\begin{equation}\label{ueqdustcls1}
		\ddot{U} + \dot{U}\left( \frac{1}{x^0} - 3\dot{P}\right) = 0
	\end{equation}
	This can be integrated to find,
	\begin{equation}\label{udefdustcls1}
		U = F_1\int \frac{\ex^{3P}}{x^0} \df x^0 + F_2
	\end{equation}
	where, $F_1,F_2$ are arbitrary functions of the coordinate $x^1$ that arise due to integration. Using equation \eqref{pdefdustcls1}, the integration in the above equation can be solved to get,
	\begin{equation}\label{pintdustcls1}
		\int \frac{\ex^{3P}}{x^0} \df x^0 = \int \frac{2 \ex^{2P}}{1 + (k+\B)\ex^{2P}}\df(\ex^P)\\ = \frac{2}{(k+\B)}\left\{ \ex^P - \frac{\tan^{-1}\left( \sqrt{k+\B}\ \ex^P\right)}{\sqrt{k+\B}} \right\} + C_2
	\end{equation}
	where, $C_2$ is a constant of integration. Using this, we get,
	\begin{equation}\label{qdefdustcls1}
		\ex^Q = F_3 - \ex^{-P}\left\{ \frac{F_3\tan^{-1}\left( \sqrt{k+\B}\ \ex^P\right)}{\sqrt{k+\B}} - F_4 \right\}
	\end{equation}
	where, $F_3(x^1) = \frac{2F_1}{k+\B}$ and $F_4(x^1) = C_2F_1 + F_2$. 
	
	Using equation \eqref{pdefdustcls1} in \eqref{mefe1}, we get the equation for the energy density,
	\begin{equation}\label{rhodefdustcls1}
		\frac{2\ex^{-2P}}{x^0}\left( \dot{P}+\dot{Q} \right) 
	\end{equation}
	
	If $F_3 = 0$ (i.e., $F_1 = 0$), we recover the vacuum solutions ($\ex^P~=~\ex^{-Q}$). Assuming that $F_3\ne 0$, we can rescale the coordinate $x^1$ such that $F_3 =1 $. Then the line element in \eqref{g3metric} takes the form,
	\begin{multline}\label{metricdustcls1}
		\df s^2 = -\frac{x^0}{C_1 - (k+\B)x^0} (\df x^0)^2 + \left[ 1 - \ex^{-P}\left\{ \frac{\tan^{-1}\left( \sqrt{k+\B}\ \ex^P\right)}{\sqrt{k+\B}} - F_4 \right\} \right]^2 (\df x^1)^2\\ + (x^0)^2 \left[ (\df x^2)^2 + S^2(x^2,k)(\df x^3)^2 \right]
	\end{multline}
	
	This reduces to the solution in general relativity when the back-reaction term, $\B$, is zero. In general relativity, this subclass of solutions contains generalisations of the Kantowski-Sachs solution (for $k=1$) \cite{kant,stephani} and Bianchi Type-I solution (for $k=0$). It would be interesting to see if the same is true for the MG analogues of these solutions. Given that the back-reaction term in MG couples with the geometry in a non-trivial manner, we expect these limits to either change or even not work at all.
	
	\subsection{Solutions with $R^\prime \ne 0$}
	\un{\textit{When $R^\prime \ne 0$}}, we can rescale the coordinate $x^0$ such that $G_{00} = 1$. This is equivalent to taking $P=0$ and then defining the `dots' to be the differentiation with respect to this new rescaled coordinate. Then, we can integrate equation \eqref{mefe2} (or \eqref{mefe3}) to get,
	\begin{equation}\label{qltb}
		Q = \ln (R^\prime) + \ln F_1
	\end{equation}
	where, $F_1(x^1)$ is an arbitrary function depending on the boundary conditions.
	
	By conveniently redefining the function $F_1$ as, $F_1(x^1)~\equiv~\frac{1}{\sqrt{(k+\B) - \varepsilon f^2}}$, we can write,
	\begin{equation}\label{qdefltb}
		\ex^{2Q} = \frac{{R^\prime}^2}{(k+\B) - \varepsilon f^2(x^1)}
	\end{equation}
	The function $f$ is an arbitrary function of the coordinate $x^1$ and the variable, $\varepsilon = (0,\pm 1)$, is to be chosen such that $\ex^{2Q}$ remains positive. This redefinition makes the other dynamical equations simpler. These expressions are valid for both static and non-static cases.
	
	In the static case, all the terms with dots go to zero. Then, equation \eqref{mefe4} implies that $\varepsilon f^2 = 0$. When used in equation \eqref{mefe1}, we get, $\rho = 0$. Therefore, the static solutions in the case of a pressureless fluid invariably reduce to vacuum solutions which, in the case of spherical symmetry ($k=1$), leads to nothing but the Schwarzschild solution in the Lema\^itre-Novikov \cite{stephani,lemaitre1,novikov} coordinates.
	
	In the non-static case, using equations \eqref{qltb} and \eqref{qdefltb} in \eqref{mefe1} gives,
	\begin{equation}\label{rhodefltb}
		\frac{{\dot{R}}^2 + \varepsilon f^2}{R^2} + \frac{2 \dot{R}\dot{R}^\prime + 2 \varepsilon f f^\prime}{R R^\prime} = 8\pi \rho
	\end{equation}
	Then, equation \eqref{mefe4} gives the following differential equation for $R$,
	\begin{equation}\label{reqltb}
		2R\ddot{R} + {\dot{R}}^2 = -\varepsilon f^2
	\end{equation}
	Performing an integration once, this equation gives,
	\begin{equation}\label{reqltbint}
		{\dot{R}}^2 - \frac{2m(x^1)}{R} = -\varepsilon f^2
	\end{equation}
	where, $m(x^1)$ is another arbitrary function coming from integration. This equation can be completely integrated for both the cases, $\varepsilon=0,\ \varepsilon \ne 0$ (see chapter 15 in \cite{stephani} for a complete analysis). Using equation \eqref{reqltb} and its differentiation, the other two field equations \eqref{mefe5} are identically satisfied.
	
	Using equation \eqref{reqltbint} to substitute values of $\varepsilon f^2$ and $\varepsilon f f^\prime$ in equation \eqref{reqltb}, we get\footnote{Note that equation \eqref{rhoeqmassfunltb} can be rearranged to give, \[  2m(x^1) = 8\pi \int \rho R^2 R^\prime\df x^1\] This equation resembles that for a mass function and hence justifies the labelling of the function arising from the integration as `$m$'.},
	\begin{equation}\label{rhoeqmassfunltb}
		\frac{2m^\prime}{R^2R^\prime} = 8\pi \rho
	\end{equation}
	Similarly, if we use equation \eqref{rhodefltb} to do the same, we get another equation for the density, in terms of the function $R$ only,
	\begin{equation}\label{rhoeqR}
		\frac{\ddot{R}^\prime}{R^\prime} + \frac{2 \ddot{R}}{R} = -4 \pi\rho
	\end{equation}
	Using equation \eqref{qdefltb}, the line element in equation \eqref{g3metric} takes the form,
	\begin{equation}\label{dustmetric}
		\df s^2 = -(\df x^0)^2 + \frac{{R^\prime}^2}{(k+\B) - \varepsilon f^2}(\df x^1)^2 + R^2\left[ (\df x^2)^2 + S^2(x^2,k)(\df x^3)^2 \right]
	\end{equation}
	Therefore, the dust solution is modified to have a back-reaction term. The usual dust solution can be obtained by putting the back-reaction term to be zero.
	
	\subsubsection{The MG-LTB Solution} \label{sec-mgltb}
	Our analysis until this point is done for both plane and spherical symmetry in a combined way. Although there has been some interest in the plane symmetric cosmological models in the past, none of these models has stood the tests of precision cosmology as well as the spatially homogeneous and isotropic FLRW model. The only exciting alternative to the FLRW model is the Lema\^itre-Tolman-Bondi (LTB) \cite{lemaitre,tolman,bondi} model. The LTB model is a subcase (for $R^\prime \ne 0$) of the spherically symmetric dust solutions to the Einstein field equations. It is only radially inhomogeneous (due to spherical symmetry) with the spatial hypersurfaces spherically symmetric about a central worldline. In this section, we will look at the LTB model in the framework of macroscopic gravity -- the MG-LTB model. To do this, we choose spherical coordinates $ (t,r,\theta,\phi) $ and $k=1$ in equations \eqref{qdefltb} and \eqref{dustmetric}. The line element then looks like,
	\begin{equation}\label{mgltbmetric}
		\df s^2 = -\df t^2 + \frac{{R^\prime}^2}{(1+\B) - \varepsilon f^2}\df r^2 + R^2\left[ \df \theta^2 + \sin^2 \theta\df \phi^2 \right]
	\end{equation}
	This is a modified form of LTB metric \cite{kras,molina} where there is a back-reaction term in the denominator of the radial metric coefficient. 
	
	The dynamical equations governing this model would be given by equations \eqref{rhodefltb} - \eqref{rhoeqR}. Comparing them to the Friedman equations, we can define local cosmological parameters within the LTB models \cite{enqvist1,enqvist2,bellido,alnes},
	\begin{subequations}\label{denparaltb}
		\begin{align}
			H(t,r) &= \frac{\dot{R}(t,r)}{R(t,r)} \\
			\Omega_{m_0}(r) &= \frac{2 m(r)}{H_0^2(r) R_0^3(r)} \\
			\Omega_{c_0}(r) &= \frac{-\varepsilon f^2}{R_0^2(r)H_0^2(r)}
		\end{align}
	\end{subequations}
	where, $H,\ \Omega_{m_0},\ \Omega_{c_0}$ are the local Hubble parameter, matter density parameter and curvature density parameter, respectively. The functions, $H_0(r)~\equiv~H(t_0,r)$ and $R_0(r)~\equiv~R(t_0,r)$, are the boundary values at the present time, $t_0$. Using equation \eqref{denparaltb} in \eqref{reqltbint}, we can get an expression for the expansion history in LTB models,
	\begin{equation}\label{exphistltb}
		H^2(t,r) = H_0^2(r) \left[ \Omega_{m_0}(r) \left( \frac{R_0}{R}\right)^3 \right. \\+\left. \Omega_{c_0}(r) \left(\frac{R_0}{R}\right)^2 \right]
	\end{equation}
	Integrating this equation, we get an equation for the age of the LTB universe,
	\begin{equation}\label{ageltb}
		t_0 - t = \frac{1}{H_0}\int_{R}^{R_0} \frac{\df R}{\left( \Omega_{m_0} R_0^2 R^{-1} + \Omega_{c_0} R_0^2 \right)^{\hf}}
	\end{equation}
	
	The next thing to look at, from the point of view of observational cosmology, would be light propagation. For a radial null geodesic, $\df \theta = 0 = \df \phi$, and from equation \eqref{mgltbmetric}, we get a constraint equation for the light rays \cite{enqvist1,enqvist2,bellido,alnes},
	\begin{equation}\label{radnullgeoltb}
		\frac{\df t}{\df \lambda} = -\frac{\df r}{\df \lambda} \frac{R^\prime(t,r)}{ \sqrt{1+\B - \varepsilon f^2} }
	\end{equation}
	where, $\lambda$ is an affine parameter and the negative sign tells that we are dealing with incoming light rays. Let $t(\lambda)$ and $\tau(\lambda)$ be two solutions to the equation above. Then, $t+\tau$ would be a solution too. Then, we can write,
	\begin{subequations}
		\begin{align}
			\frac{\df t}{\df \lambda} &= -\frac{\df r}{\df \lambda} \frac{R^\prime(t,r)}{ \sqrt{1+\B - \varepsilon f^2} }\label{nullgeoltb1}\\	
			\frac{\df (t+\tau)}{\df \lambda} &= -\frac{\df r}{\df \lambda} \frac{R^\prime(t,r)}{ \sqrt{1+\B - \varepsilon f^2} } + \frac{\df \tau(\lambda)}{\df \lambda} \label{nullgeoltb2}\\	
			\frac{\df (t+\tau)}{\df \lambda} &= -\frac{\df r}{\df \lambda} \frac{R^\prime(t,r)+\dot{R}^\prime(t,r)\tau}{ \sqrt{1+\B - \varepsilon f^2} } \label{nullgeoltb3}
		\end{align}
	\end{subequations}
	Using equations \eqref{nullgeoltb2} and \eqref{nullgeoltb3}, we get,
	\begin{equation}\label{nullgeoltb4}
		\frac{1}{\tau}\frac{\df \tau}{\df \lambda} = -\frac{\df r}{\df \lambda} \frac{\dot{R}^\prime(t,r)}{ \sqrt{1+\B - \varepsilon f^2} }
	\end{equation}
	Defining the redshift, $z$, as, $1+z~\equiv~\frac{\tau(0)}{\tau(\lambda)}$, we can write,
	\begin{equation}\label{redlamrel}
		\frac{\df z}{\df \lambda} = (1+z)\frac{\df r}{\df \lambda} \frac{\dot{R}^\prime(t,r)}{ \sqrt{1+\B - \varepsilon f^2} }
	\end{equation}
	This gives us,
	\begin{equation}\label{refrrel}
		\frac{\df r}{\df z}  = \frac{1}{1+z} \frac{\sqrt{1+\B - \varepsilon f^2}}{\dot{R}^\prime(t,r)}
	\end{equation}
	Further, using equations \eqref{radnullgeoltb} and \eqref{redlamrel}, we get,
	\begin{equation}\label{redtrel}
		\frac{\df t}{\df z} = -\frac{1}{1+z} \frac{R^\prime(t,r)}{\dot{R}^\prime(t,r)}
	\end{equation}
	It is, then, straightforward to calculate the comoving, angular diameter and luminosity distances, which are given by \cite{enqvist1,enqvist2,bellido,alnes}, 
	\begin{subequations}\label{distmgltb}
		\begin{align}
			d_C(z) &= (1+z) R(t(z),r(z)) \label{codistmgltb} \\
			d_A(z) &=  R(t(z),r(z)) \label{angdistmgltb}\\
			d_L(z) &= (1+z)^2 R(t(z),r(z)) \label{lumdistmgltb}
		\end{align}
	\end{subequations}
	These equations tell us that the back-reaction term will modify the redshift dependence of the coordinates, $t$ and $r$, and hence, the distance calculations. 
	
	Looking at equations \eqref{denparaltb} - \eqref{ageltb}, we can see that the equations governing the MG-LTB solution are the same as the ones for the GR-LTB solution. But the geometry of MG-LTB model (equation \eqref{mgltbmetric}) is not the same. This points to an important feature of models within macroscopic gravity, which is that the back-reaction (the connection correlation) can be set in such a way that it only affects either the averaged geometry or the averaged evolution but not both. For example, we could have chosen the integration function $F_1$ in equation \eqref{qltb} to be simply $\frac{1}{\sqrt{k-\varepsilon f^2}}$, and then, equation \eqref{reqltbint} would have had a back-reaction term, $\frac{\B}{R^2}$, in it. The FLRW limit of the LTB model is also affected in a similar fashion -- either we end up with a geometry that is not FLRW but equation \eqref{reqltbint} reduces to the usual Friedmann equation or we have an FLRW geometry that evolves differently. In the next section, we will see that something similar happens in the case of the homogeneous perfect fluid solutions as well.

	\section{Perfect Fluid} \label{subsec-perfectfluid}
	In the case of perfect fluid, the dynamical equations are the same as equations \eqref{conseq} and \eqref{genmefe}. Perfect fluid solutions are more complicated to solve analytically. In GR, the additional condition due to the isotropy of the pressure looks like, ${E^1}_1 - {E^3}_3 = 0$. However, in MG, this condition gets modified due to the anisotropy in the effective pressure due to the back-reaction and we have, ${E^1}_1 - {E^3}_3 = \frac{\B}{R^2}$. 
	
	\subsection{Static Solutions}
	In the static case, all the terms with `dots' in equations \eqref{conseq} and \eqref{genmefe} go to zero and we can assume $R=x^1$ \cite{stephani,goenner}. Then, equation \eqref{mefe1} becomes,
	\begin{equation}\label{pfeq1}
		{E^0}_0 = \left[ x^1(k+\B - \ex^{-2Q})  \right]^\prime = 8\pi \rho (x^1)^2
	\end{equation}
	We can easily integrate this equation to get,
	\begin{equation}\label{qdefspf}
		\ex^{2Q} = \left[ (k+\B) - \frac{2m(x^1)}{x^1}\right]^{-1}
	\end{equation}
	where, $m(x^1)$ is a mass function defined as,
	\begin{equation}\label{massfunc}
		2m(x^1) = 8\pi\int \rho(x^1) (x^1)^2 \df x^1
	\end{equation}
	The condition due to isotropy, ${E^1}_1 - {E^3}_3 = \frac{\B}{R^2}$, further gives,
	\begin{equation}\label{isocond}
		P^\dprime + (P^\prime)^2 - P^\prime Q^\prime \\=  \frac{1}{(x^1)^2}\left[ 1 + x^1(P^\prime + Q^\prime) - (k+\B)\ex^{2Q} \right] 
	\end{equation}
	Using equations \eqref{qdefspf}, \eqref{massfunc} and \eqref{isocond} in \eqref{mefe5}, we get,
	\begin{equation}\label{eq-P}
		2x^1\left[(k+\B)x^1 - 2m\right]P^\prime = 8\pi p (x^1)^3 + 2m
	\end{equation}
	Now, using equation \eqref{conseq2} to eliminate $P^\prime$ from the equation above, we get,
	\begin{equation}\label{rhopcond}
		2x^1\left[(k+\B)x^1 - 2m\right]p^\prime\\ = -(\rho + p)\left[8\pi p (x^1)^3 + 2m\right]
	\end{equation}
	This is the MG analogue of the Tolman-Oppenheimer-Volkoff (TOV) equation \cite{tolman1,opp}.
	
	The analysis until here is quite general. However, as in the case of vacuum solutions, we are interested in spherically symmetric solutions. For that, we use spherical coordinates $(t,r,\theta,\phi)$ and $k=1$ in equations \eqref{pfeq1}-\eqref{rhopcond}. We will look, in detail, at two static spherically symmetric perfect fluid solutions -- the Schwarzschild interior solution \cite{schw1} (constant density) and the Tolman VII solution \cite{tolman1} (variable density).
	
	\subsubsection{MG-Schwarzschild Interior Solution}
	The Schwarzschild interior solution is, perhaps, the best known static spherically symmetric perfect fluid solution in GR. The MG analogue of this solution has been presented in \cite{hoogenss}. It is characterised by a constant energy density $ (\rho~=~\bar{\rho}~=~{\rm constant}) $ inside a boundary, say, $r~\le~r_b$. Then, using equation \eqref{massfunc}, the mass function inside this boundary becomes,
	\begin{equation}\label{massint}
		2m(r) = 8\pi\int^r_0 \bar{\rho} r^2 \df r = \frac{8 \pi}{3} \bar{\rho} r^3
	\end{equation}
	and then, using equation \eqref{qdefspf}, we get,
	\begin{equation}\label{qdefschint}
		\ex^{2Q(r)} = \left[ (1+\B) - \frac{8 \pi}{3} \bar{\rho} r^2\right]^{-1}
	\end{equation}
	
	Further, using the constant density term in equation \eqref{conseq2}, we can integrate to get,
	\begin{equation}\label{press}
		p = C_1\ex^{-P} - \bar{\rho}
	\end{equation}
	Then, using this equation in \eqref{eq-P}, we can integrate to find,
	\begin{equation}\label{pdefschint}
		\ex^{P} = \frac{3}{2\bar{\rho}}\left( C_1 - C_2 \ex^{-Q}\right) \\= \frac{3}{2\bar{\rho}}\left[C_1 - C_2\left\{ (1+\B) - \frac{8 \pi}{3} \bar{\rho} r^2\right\}^{\hf}\right]
	\end{equation}
	Putting this equation back in the expression for pressure above \eqref{press}, we get,
	\begin{equation}\label{pressschint}
		p = \frac{\bar{\rho}}{3} \left(\frac{3\ C_2 \ex^{-Q(r)} - C_1}{C_1 - C_2\ex^{-Q(r)}}\right) = \frac{\bar{\rho}}{3} \frac{3\ C_2 \left[ (1+\B) - \frac{8 \pi}{3} \bar{\rho} r^2\right]^{\hf} - C_1}{C_1 - C_2 \left[ (1+\B) - \frac{8 \pi}{3} \bar{\rho} r^2\right]^{\hf}}
	\end{equation}
	
	The values for the constants $C_1,C_2$ can be determined by matching the interior solution to the exterior solution. The expression for $\ex^{2Q}$ already matches the one for the exterior solution. Requiring that, for some boundary $r_b$, $p(r_b) = 0$, gives us, $ \frac{C_1}{C_2}~=~3\ex^{-Q(r_b)} $. Using this, the expression for pressure becomes,
	\begin{equation}\label{press-cosnt}
		p = \bar{\rho} \left(\frac{ \ex^{-Q(r)} - \ex^{-Q(r_b)} }{3\ \ex^{-Q(r_b)}  - \ex^{-Q(r)} }\right) = \bar{\rho} \frac{\left[ (1+\B) - \frac{8 \pi}{3} \bar{\rho} r^2\right]^{\hf} - \left[ (1+\B) - \frac{8 \pi}{3} \bar{\rho} r_b^2\right]^{\hf}}{3\left[ (1+\B) - \frac{8 \pi}{3} \bar{\rho} r_b^2\right]^{\hf} - \left[ (1+\B) - \frac{8 \pi}{3} \bar{\rho} r^2\right]^{\hf}}
	\end{equation} 	
	Further, we require that $\ex^{P(r_b)}\left|_{\rm int}\right.~=~\ex^{P(r_b)}\left|_{\rm ext}\right. $. Using this, we can completely specify the constants as,
	\begin{equation}\label{schwextconst}
		C_1 = \bar{\rho} \left\{ (1+\B) - \frac{8 \pi}{3} \bar{\rho} r_b^2\right\}^{\hf} \; ; \quad C_2 = \frac{\bar{\rho}}{3}
	\end{equation}
	Then, equation \eqref{pdefschint} becomes,
	\begin{equation}\label{pdef-const}
		\ex^{P} = \hf\left[3 \left\{ (1+\B) - \frac{8 \pi}{3} \bar{\rho} r_b^2\right\}^{\hf}\left\{ (1+\B) - \frac{8 \pi}{3} \bar{\rho} r^2\right\}^{\hf}\right]
	\end{equation}
	
	Therefore, the MG-Schwarzschild interior solution is characterised by a constant density $\bar{\rho}$ and a pressure given by equation \eqref{press-cosnt}. The metric coefficients are given by equations \eqref{qdefschint} and \eqref{pdef-const}. The line element for the interior solution then looks like,
	\begin{multline}\label{mgschwint}
		\df s^2 = -\frac{1}{4}\left[3 \left\{ (1+\B) - \frac{8 \pi}{3} \bar{\rho} r_b^2\right\}^{\hf} - \left\{ (1+\B) - \frac{8 \pi}{3} \bar{\rho} r^2\right\}^{\hf}\right]^2 \df t^2 \\+  \frac{1}{\left[ (1+\B) - \frac{8 \pi}{3} \bar{\rho} r^2\right]} \df r^2 + r^2\df \theta^2 + r^2 \sin^2 \theta \df \phi^2
	\end{multline} 
	
	The Schwarzschild interior and exterior solutions derived here match the ones presented in \cite{hoogenss}. However, presenting all the calculations again is justified since we are working with fewer restrictions on the connection correlation. This adds to the completeness of the analysis in this paper. In the next section, we will present an interior solution with non-constant density.	
	
	\subsubsection{MG-Tolman VII Solution}
	The obvious solution to look at, after the Schwarzschild solution, is the Tolman VII solution \cite{tolman1}.	The Tolman VII solution is characterised by the following ansatz on the energy density \cite{raghoo},
	\begin{equation}\label{rhotol}
		\rho = \rho_0 \left[ 1 - \beta \left( \frac{r}{r_b} \right)^2\right]
	\end{equation} 
	where, $\rho_0$ is the central density ($\rho_0 \equiv\rho(r=0)$), $r_b$ represents a boundary radius beyond which the solution can be considered to be the Schwarzschild exterior solution, and  $\beta$ is a dimensionless parameter that takes values between 0 and 1.
	
	Using this ansatz for the energy density, the mass function (equation \eqref{massfunc}) inside the boundary becomes,
	\begin{equation}\label{masstol7}
		2m(r) = 8\pi \rho_0 \left(\frac{r^3}{3} - \frac{\beta r^5}{5 r_b^2} \right)
	\end{equation}
	and then, using equation \eqref{qdefspf}, we get\footnote{In the original paper by Tolman \cite{tolman1}, this form for the metric coefficient was assumed and the expression for the density was derived using that.},
	\begin{equation}\label{qdeftol7}
		\ex^{2Q} = \left[ (1+\B) - 8\pi \rho_0 \left(\frac{r^2}{3} - \frac{\beta}{5 r_b^2} r^4 \right)\right]^{-1}
	\end{equation}
	We define a new variable $x\equiv\frac{r^2}{r_b^2}$. Then, the above equation can be written as,
	\begin{equation}\label{qdeftol7newvar}
		\ex^{2Q} = \left[ (1+\B) - A \left(\frac{x}{3} - \frac{\beta}{5} x^2 \right)\right]^{-1}
	\end{equation}
	where, $A=8\pi \rho_0r_b^2$, is simply a constant.
	
	Rearranging equation \eqref{isocond}, we get,
	\begin{equation}\label{isocondtol7}
		\ex^{-2Q}\left[ 1 + rP^\prime + rQ^\prime(1+rP^\prime)\right.\\\left. - r^2(P^\dprime + {P^\prime}^2)\right] = (1+\B)
	\end{equation}
	Using equation \eqref{qdeftol7} and its differentiation with respect to $r$ to find the value of $ Q^\prime\ex^{-2Q} $, the above equation becomes,
	\begin{equation}
		\left[ (1+\B) - A \left(\frac{x}{3} - \frac{\beta}{5} x^2 \right)\right] r^2(P^\dprime + {P^\prime}^2) \\- \left[ (1+\B) - A  \frac{\beta}{5} x^2 \right]rP^\prime - A\frac{\beta}{5} x^2 = 0
	\end{equation}
	Then, following \cite{mehra,durgapal1}, we define a function $U(r)$, such that, $P~\equiv~\ln U$ and simplify further to get,
	\begin{equation}
		\left( \frac{5(1+\B)}{\beta A} -  \frac{5x}{3\beta} + x^2 \right) r^2 U^\dprime \\- \left( \frac{5(1+\B)}{\beta A} -  x^2 \right)rU^\prime - x^2U = 0
	\end{equation}
	Writing the differential terms to be with respect to the variable, $ x $, we get,
	\begin{equation}\label{diffeqx}
		\underbrace{\left( \frac{5(1+\B)}{\beta A} -  \frac{5x}{3\beta} + x^2 \right)}_{M} \widetilde{\widetilde{U}} + \underbrace{\left( x - \frac{5}{6\beta} \right)}_{N}\widetilde{U} - \frac{U}{4} = 0
	\end{equation}
	where `tilde' represents differentiation with respect to $x$ and we have used, $U^\prime = \frac{2r}{r_b^2}\tilde{U}$ and $U^\dprime = \frac{2r}{r_b^2}\tilde{U} + \frac{4r^2}{r_b^4}\tilde{\tilde{U}}$.
	
	Now, we define a variable $w~=~\ln (\sqrt{M} + N)$. This gives, $\tilde{w}~=~\frac{1}{\sqrt{M}}$, and, $ \tilde{w}~=~-N M\sqrt{M} $. Using this, we get, $\widetilde{U}~=~\frac{1}{\sqrt{M}}\frac{\df U}{\df w}$, and, $ \widetilde{\widetilde{U}}~=~\frac{1}{M}\frac{\df^2 U}{\df w^2}~-~\frac{N}{M\sqrt{M}}\frac{\df U}{\df w}$. Substituting these in equation \eqref{diffeqx}, we get,
	\begin{equation}\label{diffeqw}
		\frac{\df^2 U}{\df w^2} + \frac{U}{4} = 0
	\end{equation}
	This is the equation of a simple harmonic oscillator with frequency $\hf$ and has the following solution,
	\begin{equation}\label{usol}
		U = C_1 \cos \left(\hf w\right) + C_2 \sin \left(\hf w\right)
	\end{equation}
	where, $C_1,C_2$ are constants of integration. Then using the definition of the function $U$, we can write,
	\begin{equation}\label{pdeftolman7}
		\ex^{2P} = \left[ C_1 \cos \left(\hf w\right) + C_2 \sin \left(\hf w\right) \right]^2
	\end{equation} 
	where, $w~=~\ln \left\{ \frac{r^2}{r_b^2} - \frac{5}{6\beta} + \left( \frac{r^4}{r_b^4} - \frac{5r^2}{3\beta r_b^2} + \frac{5(1+\B)}{8\pi \beta\rho_0 r_b^2}\right)^{\hf} \right\} $.
	
	Substituting the expression for the mass function (equation \eqref{masstol7}) and using equation \eqref{usol} to find $P^\prime = \frac{U^\prime}{U}$ in equation \eqref{eq-P}, we get an expression for the pressure,
	\begin{equation}\label{presstol7}
		p = \frac{1}{r_b}\left(\frac{\beta\rho_0}{10 \pi}\right)^{\hf}\left\{ (1+\B) - 8\pi\rho_0  \left(\frac{r^2}{3} - \frac{\beta}{5r_b^2}r^4\right)\right\}^{\hf} \left\{ \frac{C_2 - C_1 \tan \left(\hf w\right)}{C_1 + C_2 \tan\left(\hf w\right)} \right\} - \rho_0\left(\frac{1}{3} - \frac{\beta}{5 r_b^2} r^2\right)
	\end{equation}
	The constants $C_1,C_2$ can be determined using the boundary conditions: $\ex^{2P(r_b)}\left|_{\rm tol}\right.~=~\ex^{2P(r_b)}\left|_{\rm ext}\right.$ and $p(r_b)~=~0$. The first one gives,
	\begin{subequations}\label{consteq}
		\begin{equation}\label{consteq1}
			C_1 \cos \left(\hf w_b\right) + C_2 \sin \left(\hf w_b\right)\\ = \left\{ (1+\B) - 8\pi\rho_0r_b^2  \left(\frac{1}{3} - \frac{\beta}{5}\right)\right\}^{\hf}
		\end{equation}
		\textrm{Using this, the second boundary condition gives,}
		\begin{equation}\label{consteq2}
			C_2 \cos \left(\hf w_b\right) - C_1 \sin \left(\hf w_b\right) \\= r_b\left(\frac{10\pi\rho_0}{\beta}\right)^{\hf}\left(\frac{1}{3} - \frac{\beta}{5}\right)
		\end{equation}
	\end{subequations}
	where, $w_b\equiv w(r_b)$. Solving equations \eqref{consteq}, we get,
	\begin{subequations}\label{consttol7}
		\begin{equation}\label{consttol71}
			C_1 = \left\{ (1+\B) - 8\pi\rho_0r_b^2  \left(\frac{1}{3} - \frac{\beta}{5}\right)\right\}^{\hf} \cos \left(\hf w_b\right) \\  - r_b\left(\frac{10\pi\rho_0}{\beta}\right)^{\hf}\left(\frac{1}{3} - \frac{\beta}{5}\right) \sin \left(\hf w_b\right)
		\end{equation}
		\begin{equation}\label{consttol72}
			C_2 = \left\{ (1+\B) - 8\pi\rho_0r_b^2  \left(\frac{1}{3} - \frac{\beta}{5}\right)\right\}^{\hf} \sin \left(\hf w_b\right) \\  +  r_b\left(\frac{10\pi\rho_0}{\beta}\right)^{\hf}\left(\frac{1}{3} - \frac{\beta}{5}\right) \cos \left(\hf w_b\right)
		\end{equation}
	\end{subequations}
	where, $ w_b = \ln \left\{ 1 - \frac{5}{6\beta} + \left( 1 - \frac{5}{3\beta} + \frac{5(1+\B)}{8\pi\beta\rho_0 r_b^2}\right)^{\hf}\right\} $.
	
	Using these expressions for the constants and simplifying even further, equations \eqref{pdeftolman7} and \eqref{presstol7} become,
	\begin{multline}\label{pdeftol7-const}
		\ex^{2P} = \left[\left\{ (1+\B) - 8\pi\rho_0r_b^2  \left(\frac{1}{3} - \frac{\beta}{5}\right)\right\}^{\hf} \cos \left\{\ln \left( \frac{\frac{r^2}{r_b^2} - \frac{5}{6\beta} + \left( \frac{r^4}{r_b^2} - \frac{5r^2}{3\beta r_b^2} + \frac{5(1+\B)}{8\pi \beta\rho_0 r_b^2}\right)^{\hf}}{ 1 - \frac{5}{6\beta} + \left( 1 - \frac{5}{3\beta} + \frac{5(1+\B)}{8\pi\beta\rho_0 r_b^2}\right)^{\hf}}\right)^{\hf}\right\} \right.\\ \left.  + r_b\left(\frac{10\pi\rho_0}{\beta}\right)^{\hf}\left(\frac{1}{3} - \frac{\beta}{5}\right) \sin \left\{\ln \left( \frac{\frac{r^2}{r_b^2} - \frac{5}{6\beta} + \left( \frac{r^4}{r_b^2} - \frac{5r^2}{3\beta r_b^2} + \frac{5(1+\B)}{8\pi \beta\rho_0 r_b^2}\right)^{\hf}}{ 1 - \frac{5}{6\beta} + \left( 1 - \frac{5}{3\beta} + \frac{5(1+\B)}{8\pi\beta\rho_0 r_b^2}\right)^{\hf}}\right)^{\hf}\right\} \right]^2
	\end{multline}
	and,
	\begin{multline}\label{presstol7-const}
		p = \frac{\beta \left\{ (1+\B) - 8\pi\rho_0r_b^2  \left(\frac{1}{3} - \frac{\beta}{5}\right)\right\}^{\hf}}{10\pi r_b^2 \left(\frac{1}{3} - \frac{\beta}{5}\right)}\left\{ (1+\B) - 8\pi\rho_0  \left(\frac{r^2}{3} - \frac{\beta}{5r_b^2}r^4\right)\right\}^{\hf}\\ \times \left[ \frac{1 +  r_b\left(\frac{10\pi\rho_0}{\beta}\right)^{\hf}\left(\frac{1}{3} - \frac{\beta}{5}\right)\tan \left\{\ln \left( \frac{\frac{r^2}{r_b^2} - \frac{5}{6\beta} + \left( \frac{r^4}{r_b^2} - \frac{5r^2}{3\beta r_b^2} + \frac{5(1+\B)}{8\pi \beta\rho_0 r_b^2}\right)^{\hf}}{ 1 - \frac{5}{6\beta} + \left( 1 - \frac{5}{3\beta} + \frac{5(1+\B)}{8\pi\beta\rho_0 r_b^2}\right)^{\hf}}\right)^{\hf}\right\}}{1 - \left\{ (1+\B) - 8\pi\rho_0r_b^2  \left(\frac{1}{3} - \frac{\beta}{5}\right)\right\}^{\hf} \tan\left\{\ln \left( \frac{\frac{r^2}{r_b^2} - \frac{5}{6\beta} + \left( \frac{r^4}{r_b^2} - \frac{5r^2}{3\beta r_b^2} + \frac{5(1+\B)}{8\pi \beta\rho_0 r_b^2}\right)^{\hf}}{ 1 - \frac{5}{6\beta} + \left( 1 - \frac{5}{3\beta} + \frac{5(1+\B)}{8\pi\beta\rho_0 r_b^2}\right)^{\hf}}\right)^{\hf}\right\}} \right]\\ - \rho_0\left(\frac{1}{3} - \frac{\beta}{5 r_b^2} r^2\right)
	\end{multline}
	The line element for this solution then looks like,
	\begin{multline}\label{mgtol7}
		\df s^2 = \left[\left\{ (1+\B) - 8\pi\rho_0r_b^2  \left(\frac{1}{3} - \frac{\beta}{5}\right)\right\}^{\hf} \cos \left\{\ln \left( \frac{\frac{r^2}{r_b^2} - \frac{5}{6\beta} + \left( \frac{r^4}{r_b^2} - \frac{5r^2}{3\beta r_b^2} + \frac{5(1+\B)}{8\pi \beta\rho_0 r_b^2}\right)^{\hf}}{ 1 - \frac{5}{6\beta} + \left( 1 - \frac{5}{3\beta} + \frac{5(1+\B)}{8\pi\beta\rho_0 r_b^2}\right)^{\hf}}\right)^{\hf}\right\} \right.\\ \left.  + r_b\left(\frac{10\pi\rho_0}{\beta}\right)^{\hf}\left(\frac{1}{3} - \frac{\beta}{5}\right) \sin \left\{\ln \left( \frac{\frac{r^2}{r_b^2} - \frac{5}{6\beta} + \left( \frac{r^4}{r_b^2} - \frac{5r^2}{3\beta r_b^2} + \frac{5(1+\B)}{8\pi \beta\rho_0 r_b^2}\right)^{\hf}}{ 1 - \frac{5}{6\beta} + \left( 1 - \frac{5}{3\beta} + \frac{5(1+\B)}{8\pi\beta\rho_0 r_b^2}\right)^{\hf}}\right)^{\hf}\right\} \right]^2 \df t^2 \\ +  \frac{1}{\left[ (1+\B) - 8\pi \rho_0 \left(\frac{r^2}{3} - \frac{\beta}{5 r_b^2} r^4 \right)\right]} \df r^2 + r^2\df \theta^2 + r^2 \sin^2 \theta \df \phi^2
	\end{multline} 
	The MG-Tolman VII solution is completely specified by equations \eqref{rhotol}, \eqref{qdeftol7}, \eqref{pdeftol7-const}, \eqref{presstol7-const} and \eqref{mgtol7}. It reduces to the usual Tolman VII solution in \cite{mehra} if we put the back-reaction term to be zero.
	
	Note that the static perfect fluid solutions presented here are not the most general for their assumed form of density, since we took the constant of integration in the mass function to be zero. In the more general solutions, there would be a term $\propto\frac{1}{r}$ in the metric coefficient $G_{11}$ (equations \eqref{qdefschint} and \eqref{qdeftol7}) which would lead to a singularity at $r=0$. We only mention this for the sake of completeness and these cases are beyond the scope of this paper.
	
	\subsection{Non-Static Solutions}
	The non-static plane or spherically symmetric perfect fluid solutions are usually classified according to the properties of the kinematic variables \cite{stephani,herlt},
	\begin{subequations}\label{kinvar}
		\begin{align}	
			\Theta &= \ex^{-P}\left(\dot{Q} + 2\frac{\dot{R}}{R}\right) \label{expansion}\\
			\dot{u}_\alpha &= \left[ 0,P^\prime,0,0 \right] \label{acceleration}\\			
			\sigma^0_0 &= 0 \; ;\; -\hf\sigma^1_1 = \sigma^2_2 = \sigma^3_3 = \frac{1}{3}\ex^{-P}\left(\frac{\dot{R}}{R} - \dot{Q}\right)\label{shear}\\
			\intertext{And due to the symmetry here, we have,}
			\omega_{\alpha\beta} & = 0 
		\end{align}
	\end{subequations}	
	where, $ \Theta, \dot{u}, \sigma $ and $\omega$ are the 4-velocity's expansion, acceleration, shear and rotation (or vorticity), respectively \cite{stephani,poisson}. A wide variety of solutions exist with different assumptions on these variables.
	
	\subsubsection{Solutions with $\Theta= 0$; ${\sigma^\alpha}_\beta = 0$}
	The vanishing shear and expansion, through equations \eqref{expansion} and \eqref{shear}, imply that $\dot{Q}~= 0~=~\dot{R}$. Then, equation \eqref{conseq1} implies that the energy density does not depend on $x^0$ (given that $\rho~+~p~\ne~0$). This will modify the field equations such that no derivative with respect to $x^0$ is present there. This is the same as the static case. However, one can still have $\dot{P}~\ne~0$. Then, for the pressure, we have $\dot{p}~\ne~0$, if $\dot{P}^\prime~\ne~0$. Therefore, these solutions are either static or can be generated from static solutions \cite{stephani}.
	
	\subsubsection{Solutions with $\Theta\ne 0$; ${\sigma^\alpha}_\beta = 0$}
	The most interesting and well studied case is that of expanding, shear free solutions. We will look at this case in some detail since it has the FLRW solution as a subcase. Vanishing shear, through equation \eqref{shear}, implies $\dot{Q} = \frac{\dot{R}}{R}$. This can be integrated to give,
	\begin{equation}\label{reqnspf}
		R = X(x^1)\ex^Q
	\end{equation}
	where, $X$ is an arbitrary function arising from integration. Using this in equation \eqref{mefe2} and then integrating, we get,
	\begin{equation}\label{peqnspf}
		\ex^P = \dot{Q}\ex^{-Y(x^0)}
	\end{equation}
	where, $Y$ is another arbitrary function coming from integration. Then, using this and equation \eqref{reqnspf}, the expansion scalar becomes, $\Theta = 3\ex^Y$.
	
	Using equations \eqref{reqnspf} and \eqref{peqnspf} in equations \eqref{mefe1} and \eqref{mefe4}, the energy density and pressure can be written as,
	\begin{equation}\label{dennspf}
		3\ex^{2Y} - \frac{\ex^{-2Q}}{X^2} \left(2XX^\dprime + {X^\prime}^2 + 2X^2Q^\dprime \right.\\\left.+ 4 X X^\prime Q^\prime + X^2 {Q^\prime}^2 + k+\B\right) = 8\pi \rho
	\end{equation}  
	and,
	\begin{equation}\label{pressnspf}
		\frac{\ex^{-3Q}}{\dot{Q}X^2} \partial_0\left[\ex^Q\left\{ \left(X^\prime + XQ^\prime\right)^2 - (k+\B)\right.\right.\\\left.\left. - \ex^{2(Q+Y)}\right\}\right]= 8\pi p
	\end{equation}
	The isotropy condition, ${E^1}_1 - {E^3}_3 = \frac{\B}{R^2}$, gives,
	\begin{equation}\label{isocondnspf}
		\partial_0 \left[\ex^Q \left\{ {X^\prime}^2 - X\left(X^\dprime + X^\prime Q^\prime\right)\right.\right.\\\left.\left. + X^2\left({Q^\prime}^2 - Q^\dprime\right) - (k+\B) \right\} \right] = 0
	\end{equation}
	Further, the metric becomes,
	\begin{equation}\label{metricnspf}
		\df s^2 = -\dot{Q}^2 \ex^{-2Y} (\df x^0)^2 + \ex^{2Q} (\df x^1)^2 \\+ \ex^{2Q} X^2 \left[ (\df x^2)^2 + S^2(x^2,k) (\df x^2)^2 \right]
	\end{equation}
	
	Equation \eqref{isocondnspf} can be integrated and then further solved to produce several classes of solutions \cite{stephani}, especially in the case of spherical symmetry. The one that is of interest to us is the spherically symmetric solution $ ((k=1) ;\ x^\mu=(t,r,\theta,\phi)) $ with a homogeneous distribution of matter density and pressure  $(\rho~\equiv~\rho(t) ;\ p~\equiv~p(t))$. The homogeneity of the pressure, through equation \eqref{conseq2}, implies $P^\prime~=~0$. As explained in section \ref{subsec-dust}, this allows one to put $P=0$ and write the metric coefficient $Q$ as in equation \eqref{qdefltb}. Further equation \eqref{peqnspf}, then, implies $\dot{Q}^\prime~=~0$. Using this, we can write,
	\begin{equation}\label{qcond}
		Q~=~Q_1(t) + Q_2(r)
	\end{equation}
	Using this in equation \eqref{reqnspf}, we get,
	\begin{equation}\label{rprimecond}
		R^\prime = Z^\prime a(t)
	\end{equation}	
	where, $Z(r)=X\ex^{Q_2(r)}$ and $a(t) = e^{Q_1(t)}$.
	
	Then, using equations \eqref{qdefltb} and \eqref{rprimecond}, the metric in equation \eqref{metricnspf} (for the spherically symmetric case) becomes,
	\begin{equation}\label{metricnspf1}
		\df s^2 = -\df t^2 + \frac{{Z^\prime}^2 a^2}{F^2(r)} \df r^2 + Z^2 a^2 \left(\df \theta^2 + \sin^2\theta\df \phi^2\right)
	\end{equation}
	where, the function $F$ is an arbitrary function that comes from integration. Now, since $Z,Z^\prime$ are functions of only $r$, we can perform a coordinate transformation $r=Z(r)$ such that the above metric becomes,
	\begin{equation}\label{flrwmetric}
		\df s^2 = -\df t^2 + a^2\left(\frac{\df r^2}{F^2(r)}  + r^2\df \theta^2 + r^2\sin^2\theta\df \phi^2\right)
	\end{equation}
	This is nothing but the spatially homogeneous and isotropic FLRW metric.
	
	The function $F$ can be (conveniently) redefined such that $F^2(r)~=~(1~+~\B)~-~Kr^2$. Given this redefinition, the MG field equations reduce to the usual Friedmann equations. On the other hand, if $F^2$ is taken to be simply $1~-~Kr^2$, we recover the usual FLRW geometry, but now, its evolution is affected by the back-reaction term. This highlights the obvious, albeit important, point that we discussed briefly in the case of LTB solution, which is, in MG, the FLRW geometry evolves differently than in GR and the geometry that evolves like a FLRW geometry is different from FLRW.  
	
	\subsubsection{Solutions with ${\sigma^\alpha}_\beta\ne 0$; $\dot{u}^\alpha=0$}
	The vanishing acceleration, through equation \eqref{acceleration}, implies that $P^\prime~=~0$. Then, equation \eqref{conseq2} would mean that $p^\prime~=~0$. Further, equation \eqref{mefe2} or \eqref{mefe3} gives,
	\begin{equation}\label{inteqpfshear}
		\dot{R}^\prime~=~\dot{Q}R^\prime
	\end{equation}
	Since the integration of this equation depends on what kind of a function $R$ is, the solutions with shear and vanishing acceleration are divided into the following three categories:
	
	\noindent\textit{i) \un{When $R=\textrm{constant}=C(\textrm{say})$}}, equation \eqref{mefe1} and \eqref{mefe4} gives us,
	\begin{equation}\label{rhopdefpfshear1}
		-8\pi \rho = \frac{k+\B}{C^2} = 8\pi p
	\end{equation}
	Using this, equation \eqref{mefe5} can be solved to find $Q$. Since $P~\equiv~P(x^0)$, the time coordinate can be rescaled such that $P=0$ (similar to what we had in the case of dust). The line element, in this case, looks like,
	\begin{equation}\label{metricpfshear1}
		\df s^2 = -(\df x^0)^2 + S^2\left\{\frac{x^0}{C},-(k+\B)\right\}(\df x^1)^2 \\+ C^2\left[ (\df x^2)^2 + S^2(x^2,k)(\df x^3)^2 \right]
	\end{equation}
	
	\noindent\textit{ii) \un{When $ R~\ne~\textrm{constant} $; $ R^\prime~=~0 $}}, one can choose $R~=~x^0$. The field equations then give,
	\begin{subequations}\label{feqpfshear}
		\begin{align}
			\frac{\ex^{-2P}}{(x^0)^2}\left\{ 1 + 2x^0\dot{Q} + (k+\B)\ex^{2P} \right\} &= 8\pi \rho \label{feqpfshear1}\\
			\frac{\ex^{-2P}}{(x^0)^2}\left\{ 1 - 2x^0\dot{P} - (k+\B)\ex^{2P} \right\} &= 8\pi p \label{feqpfshear2}\\
			\frac{\ex^{-2P}}{(x^0)^2}\left\{ 1 - (x^0)^2\left( \ddot{Q} + \dot{Q}^2 -\dot{P}\dot{Q}\right) \right\} &= 4\pi (\rho+3p) \label{feqpfshear3}
		\end{align}
	\end{subequations} 
	These equations can be solved to find one of the two functions $P,Q$ by prescribing the other one. We know that due to zero acceleration, $P^\prime~=~0$. However, it should be noted that, even though there is no $Q^\prime$ in the field equations, it is not zero in general.\\
	
	\noindent\textit{iii) \un{When $R^\prime~\ne~0$}}, one can choose $P~=~0$. Then, this case becomes similar to the corresponding case with dust. Equation \eqref{inteqpfshear} can be integrated to find,
	\begin{equation}\label{qdefpfshear}
		\ex^{2Q} = \frac{{R^\prime}^2}{(k+\B)-\varepsilon f^2} \qquad\quad \varepsilon = 0,\pm 1
	\end{equation}
	The energy density and pressure can be determined from the field equations, which reduce to,
	\begin{subequations}\label{rhopdefpfshear2}
		\begin{align}
			\frac{{\dot{R}}^2 + \varepsilon f^2}{R^2} + \frac{2 \dot{R}\dot{R}^\prime + 2 \varepsilon f f^\prime}{R R^\prime} &= 8\pi \rho\\
			\frac{\dot{R}^2 + \varepsilon f^2}{R^2} + \frac{2\ddot{R}}{R} &= -8\pi p\\
			\frac{\ddot{R}^\prime}{R R^\prime} + \frac{2\ddot{R}}{R} &= -4\pi(\rho + 3p)
		\end{align}
	\end{subequations}
	The first equation above is the same as the one in the case of dust and the third equation comes from differentiating the second one with respect to $x^1$. The metric takes the same form as the one on equation \eqref{dustmetric}.
	
	\section{Conclusion} \label{sec-discussion}
	In this paper, we have analysed the effects of averaging on plane and spherically symmetric macroscopic geometries within the framework of macroscopic gravity. We found that the back-reaction takes a form of an anisotropic fluid and enters the field equations in the form of an additional spatial curvature. This was consistent with the findings in \cite{coleyss1,coleyss2,hoogenss} for spherically symmetric geometries. Here, we have extended the analysis further to include plane symmetric geometries as well. We categorised the solutions based on the type of source and then analysed various subcases within.
	
	Taking the source to be vacuum led to the MG-Schwarzschild solution. Similarly, for dust, the non-static solutions led to the MG-LTB solution and for a perfect fluid, we derived Schwarzschild interior and Tolman VII (static case) and some non-static solutions. Our approach in this work was different from how back-reaction has been treated in the literature in that we considered the influence of back-reaction on the geometry instead of the dynamics. Since in MG, the back-reaction modifies the field equations, it will influence the geometry and not just the dynamical evolution.  		
	
	\section*{Acknowledgements}
		AA acknowledges that a code to find exact-FLRW solutions in MG was written by Tharake Wijenayake and Mustapha Ishak \cite{tharake1}, using the computer algebra software \href{https://www.maplesoft.com}{Maple} (\url{https://www.maplesoft.com}) and the openly available package \href{https://github.com/grtensor/grtensor}{GRTensor} (\url{https://github.com/grtensor/grtensor}). All the results in this paper were obtained by modification and expansion of this code.

	\bibliography{LTB-in-MG-arxiv}
	\bibliographystyle{unsrt}
\end{document}